\documentclass[10pt,twocolumn,prd,aps,amssymb,amsmath,tightenlines,showpacs]{revtex4}
\usepackage{graphicx}

\newcommand{\cP}{\ensuremath{\mathcal{P}}}
\newcommand{\cT}{\ensuremath{\mathcal{T}}}
\newcommand{\veps}{\varepsilon}

\begin{document}

\title{Comment on ``Numerical estimates of the spectrum for anharmonic PT
symmetric potentials'' by Bowen {\it et al}}

\author{Carl M. Bender$^1$ and Stefan Boettcher${}^2$}

% \email{cmb@wustl.edu}
% \email{sboettc@emory.edu}

\affiliation{${}^1$Department of Physics, Washington University, St. Louis, MO
63130, USA\\
${}^2$Department of Physics, Emory University, Atlanta, GA 30322, USA}

\date{\today}

\begin{abstract}
The paper by Bowen, Mancini, Fessatidis, and Murawski (2012 Phys.~Scr.~{\bf 85},
065005) demonstrates in a dramatic fashion the serious difficulties that can
arise when one rushes to perform numerical studies before understanding the
physics and mathematics of the problem at hand and without understanding the
limitations of the numerical methods used. Based on their flawed numerical work,
the authors conclude that the work of Bender and Boettcher is wrong even
though it has been verified at a completely rigorous level. Unfortunately, the
numerical procedures performed and described in the paper by Bowen {\it et al}
are incorrectly applied and wrongly interpreted.
\end{abstract}


\maketitle

The key sentence in the paper by Bowen {\it et al} \cite{R1} is in Section 3
(Discussion): ``The mystery of this study is why does this numerical method give
such radically different results from those discussed by Bender.'' As we explain
in the present Comment, the simple and straightforward explanation of this
``mystery'' is that the results reported in Bender and Boettcher \cite{R2} are
correct and that the numerical work reported in the paper by Bowen {\it et al}
is incorrect and misinterpreted.

In the original paper by Bender and Boettcher \cite{R2}, the eigenvalues of the
$\cP\cT$-symmetric Hamiltonian 
\begin{equation}
H=p^2+x^2(ix)^\veps,
\label{e1}
\end{equation}
where $\veps$ is a real parameter, were studied in great detail by using both
numerical and perturbative methods. It was shown that when $\veps\geq0$, the
eigenvalues are real, positive, and discrete and that there are no complex
eigenvalues. These conclusions were subsequently verified in rigorous studies
by Dorey, Dunning, and Tateo \cite{R3} and Shin \cite{R4}.

Before one can perform numerical calculations of the eigenvalues of the
Hamiltonian (\ref{e1}), it is essential to understand what it means to extend an
eigenvalue problem into the complex domain. The first detailed work in this area
was done by Bender and Wu \cite{R5}, who examined the behavior of the
eigenvalues of the quartic anharmonic oscillator as functions of complex
coupling constant. Further studies of analytically continued eigenvalue problems
were done by Bender and Turbiner \cite{R6}. The crucial element explained in all
of these studies is that as a parameter in an eigenvalue differential equation
is varied, the Stokes wedges in which the boundary conditions on the
eigenfunctions are imposed must rotate as functions of that parameter. 

To be specific, the eigenvalues associated with the Hamiltonian (\ref{e1}) are
determined by the complex differential equation
\begin{equation}
-\psi''(x)+x^2(ix)^\veps\psi(x)=E\psi(x)
\label{e2}
\end{equation}
and a pair of $\cP\cT$-symmetric boundary conditions in the complex-$x$
plane: $\psi(x)\to0$ as $|x|\to\infty$ with ${\rm arg}\,x$ lying inside the
opening angles of the Stokes wedges.  

The details of how to find the opening angles of Stokes wedges are not given
here because detailed explanations are given in Refs.~\cite{R2,R7,R8}. We simply
state the results: When $\veps=0$, the two Stokes wedges have opening angles
of $90^\circ$ and are centered about the positive-real and negative-real axes.
As $\veps$ increases from $0$, the Stokes wedges become thinner and rotate
downward into the lower-half complex-$x$ plane. When $\veps$ reaches the
value $2$, the opening angles of the wedges have decreased to $60^\circ$ and the
wedges no longer contain the positive- and negative-real axes.

Because the eigenvalue differential equation (\ref{e2}) must be solved along a
contour in the complex-$x$ plane that terminates in the Stokes wedges, it is 
best to solve the differential equation numerically on a complex contour by
using the standard Runge-Kutta method. The Runge-Kutta procedure described in
Refs.~\cite{R2,R6} has the obvious advantage that it can be used for any real
value of $\veps$, and not just for integer $\veps$. It is also extremely
important to understand that different pairs of Stokes wedges give different
sets of eigenvalues. For example, for the quantum harmonic oscillator
Hamiltonian $H=p^2+x^2$, which corresponds to $\veps=0$ in (\ref{e1}), if Stokes
wedges containing the real axes are used, the eigenvalues are $1,\,3,\,5,\,7,\,
\ldots$, which are strictly positive. However, if Stokes wedges containing the
imaginary axes are used, the eigenvalues are $-1,\,-3,\,-5,\,-7,\,\ldots$, which
are strictly negative.

In the paper by Bowen {\it et al} a clumsy and inappropriate numerical procedure
for calculating the eigenvalues is used in which the Hamiltonian is expanded in
terms of Harmonic-oscillator basis functions. The procedure is clumsy because
the algebra becomes much too unwieldy when $\veps$ is noninteger. Thus, in
the paper by Bowen {\it et al} the procedure was limited to the very special
cases $\veps=0,\,1,\,2,\,4,\,6$.

Furthermore, while the harmonic-oscillator eigenfunctions are complete on the
real axis, {\it these basis functions are not complete in the complex plane, so
these basis functions cannot be used unless the Stokes wedges include the real
axis.}\footnote{One has exactly the same problem with the sine and cosine
functions that are used to construct Fourier series. These trigonometric
functions form a complete basis on a portion of the real axis, but these
functions are definitely not complete in the complex plane.} Thus, the numerical
calculations that were performed by Bowen {\it et al} are completely invalid and
meaningless except for the trivial case of the harmonic-oscillator ($\veps=0$)
and the more interesting and nontrivial case $\veps=1$.

Unfortunately, Bowen {\it et al} made a long sequence of wrong arguments and
completely misinterpreted the numerical results that they obtained for the
$\veps=1$ case. While it is perfectly correct to expand $H=p^2+ix^3$ in an {\it
infinite} harmonic-oscillator basis, Bowen {\it et al} then truncate the
infinite matrix representation of the Hamiltonian, which is $\cP\cT$ symmetric,
to an $N\times N$ matrix, which is no longer $\cP\cT$ symmetric. As a result,
the finite-dimensional matrix has complex eigenvalues. These complex eigenvalues
reported by Bowen {\it et al} are merely artifacts of the truncation procedure
that they used.

Bowen {\it et al} then go on to make further serious misjudgments. The
procedure of truncating an infinite matrix is {\it variational} in character.
Thus, it can only be used to compute the low-lying eigenvalues. If the procedure
is used carefully and properly, one follows the behavior of the low-lying
eigenvalues as $N$ increases and observes that at first the behavior is
irregular and complex. (The behavior of the eigenvalues is not monotone because
the $N\times N$ matrix is not Hermitian.) However, as $N$ gets larger, one
observes that the eigenvalues settle down one-by-one starting at the low-energy
end of the spectrum, and stabilize at the real values found by Bender and
Boettcher. This stabilization process is extremely slow; for very large $N$, say
100, only about half a dozen eigenvalues can be determined with any useful
accuracy. The rest of the eigenvalues of the $N\times N$ matrix bear no
resemblance to the true eigenvalues of the Hamiltonian $H=p^2+ix^3$.
Unfortunately, Bowen {\it et al} take their numerical results for the high-lying
eigenvalues seriously in their paper even though they have no numerical accuracy
whatsoever. Indeed for the $p^2+ix^3$ Hamiltonian, WKB predicts that the $n$th
eigenvalue grows like $n^{6/5}$ for large $n$, whereas the numerical work of
Bowen {\it al} gives nothing of the sort.

Before concluding, we point out that Bowen {\it et al} could have benefited
by actually reading and understanding the review paper Ref.~\cite{R7}. They
clearly did not do so because no less than five times in their six-page paper
they make the ridiculous claim that in this review paper Bender states that the
Hamiltonian $H=p^2-x^2$ has real, discrete, negative eigenvalues. In fact, this
Hamiltonian certainly does not have real negative eigenvalues, and Bender has
never made such an absurd claim in any paper.

CMB is supported by the U.S.~Department of Energy and SB is supported by
U.S.~National Science Foundation.

\end{document}